\documentclass[final, runningheads]{llncs}
\usepackage[T1]{fontenc}
\usepackage{graphicx}
\usepackage{amsmath}
\newcommand{\itadata}{\footnotesize \textsl{Workshop Scientific HPC in the pre-Exascale era (part of ITADATA2024)}}
\usepackage{fancyhdr}
\fancypagestyle{empty}{\fancyhf{}\fancyhead[C]{\itadata}
  \fancyhead[R]{\includegraphics[width=.05\textwidth]{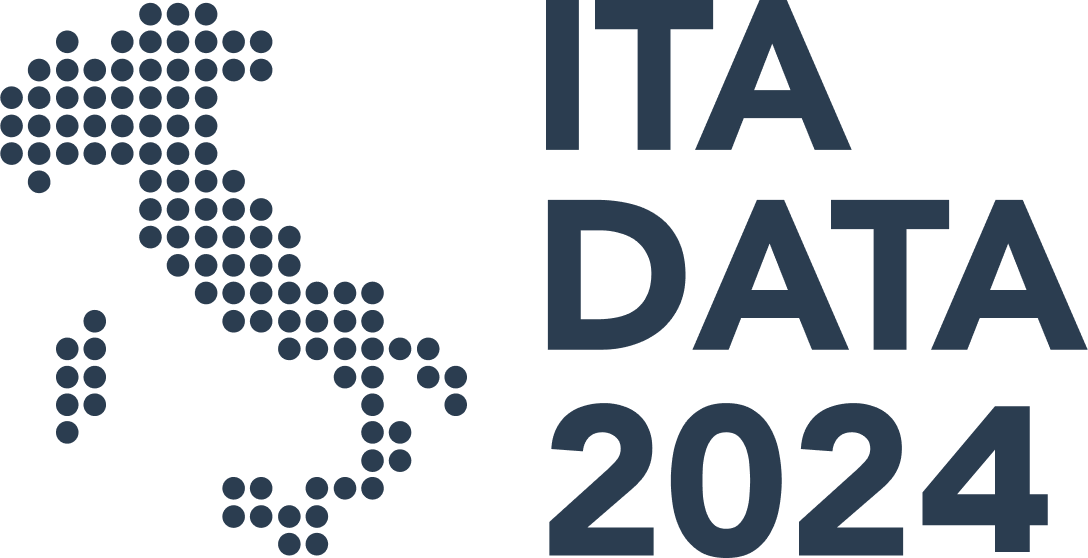}}}

\makeatletter
\begin{document}
\title{The challenge of the data in the SKA Regional Centres network}
\titlerunning{The challenge of the data in the SRCnetwork}
\author{Andrea Possenti\inst{1}\orcidID{0000-0001-5902-3731}}
\authorrunning{A. Possenti}
\institute{INAF-Osservatorio Astronomico di Cagliari, via della Scienza 5, 00947 Selargius, Italy
\email{andrea.possenti@inaf.it}}
\maketitle              \begin{abstract}
At regime, SKAO is expected to provide the researchers with an annual amount of more than 700 hundred PB of data.  The advanced analysis of all those data will take place within a network (SRCnet) of so-called SKA Regional Centres, which, under the ‘Findable, Accessible, Interoperable, and Reusable’ (FAIR) principles, will also take the responsibility for curating and archiving both the Observatory Data Products and user-generated Advanced Data Products, as well as for helping the researchers in the estimate of the needed computational effort at the proposal stage.

This contribution describes the status and the perspectives of this international network, with particular emphasis on the INAF-led Italian node which is part of that.  

\end{abstract}
\section{The SKA Observatory (SKAO) basics}

SKAO is the largest radio telescope facility ever conceived. It will be built in two very under-populated sites: {\it (1)} one (dubbed SKA-LOW) will be located in the Murchison area in Western Australia and will comprise 131072 log-periodic dipole antennas, grouped in 512 stations of 256 antennas each, spread across a maximum baseline of 74 kilometers and having a frequency range between 50 and 350 MHz;  {\it (2)} another one (dubbed SKA-MID) will be located in the Karoo area in South Africa and will include 133 dish-like antennas of 15 m in diameter, combined with other 64 antennas of 13.5 m in diameter, which are part of the already operating Meerkat radio telescope \cite{ref_article1}. The initial frequency range will be split across three observational bands, the first two running from 0.35 to 1.1 GHz and from 0.95 to 1.76 GHz, while the third covering the 4.6-15.4 GHz range. In the final configuration, additional bands will be implemented and the maximum baseline will reach about 150 kilometers. The {\it (3)} third component of SKAO, the Headquarters,  are located close to the historical, but still active, Jodrell Bank Observatory, nearby Manchester, in the UK. 

After about 3 decades during which the telescope was designed and its funding and governmental structure assessed, SKA is now in construction at the two sites, adopting a staged approach, which will lead to a first fully functional system by around the end of 2027 and 2028.

At the website \cite{ref_url1} full details can be found about the telescope configuration, the transformative science which it will deliver, the development phases, the Countries contributing to the project (16 Countries are currently involved in that as fully fledged members or at a national-coordination level or are represented as observers) and the organizational structure. 

\subsection{The key role of computing in the SKA project}

Among the experimental sciences, radio astronomy is historically one of the most demanding in terms of data acquisition, storage and computing power necessary for the full exploitation of the collected data.

In fact, radio astronomy observations require sampling and digitizing the analog signal that is received and amplified by the so-called antenna front-end. This sampling takes place both in the time domain and in the frequency domain. In many kinds of traditional observations, the band of frequencies observed is split into thousands or, sometimes, in tens of thousands of channels, and each of these channels is sampled at time intervals ranging from once per second in imaging observations until, in some time-domain observations, to more than 10000 samples per second. More recently, technology advancements allowed to also routinely collect data from an antenna up to the Nyquist frequency associated with the selected width of the observing band (the so-called baseband recording). While this is in principle the optimal choice (no information is lost in the data acquisition), it imposes very highly demanding requirements in terms of produced data: for instance, observing over a band with a width of 800 MHz (a rather common choice in modern radio astronomy) implies to sample the data at least at 1.6 GHz, translating into a data rate of 1.6 GB per second if one wants to digitize the analog signal at 8 bits. With these cadences, the flow of data coming out from each antenna of an array of antennas (or from each station of the array) can become enormous.

In the specific case of the SKAO, the design of the antennas and receivers systems is expected to produce an internal flow of data of the order of 2 PB per second, in turn pushing the requirements for the signal processing in the range of the $10^{18}$ multiply–accumulate operations (i.e. the exa-MAC range) and the associated needed high performance computing enters the ballpark of the hundreds of Petaflop/s. 

The ingestion and analysis of these data will occur in quasi-realtime at the two telescopes' sites, mainly in two large data processing machines, called CPF (Central Processing Facility) and SDP (Science Data Processor). As a result, at regime, it is expected that about 700 PB of the so-called Observatory Data Products (ODP) will be generated by the SKAO. As matter of fact, the output data rates from SKAO are extraordinary for a science experiment: of order 50-70 times higher output than other big astrophysical future facilities, like the Vera Rubin Telescope (aka LSST, operating in the optical band) and the CTAO (operating in the TeV regime).  The data rate from SKAO will be comparable to that of the High Luminosity LHC experiment at CERN.
 
As such, it is clear that in the case of SKAO the signal processing capabilities play a fundamental role in making this an unprecedented radio astronomical instrument and are at the heart of the transformative science which this new international facility promises. Indeed, in some fields of the radio astronomical research the signal processing capability of the currently available technology are the limiting factor for achieving even more exciting scientific outcomes with SKAO. If this is something to live with at the beginning, it also opens a great future for this facility, since the growing of the digital technology is expected to occur faster than any additional development of the antennas and front-ends. Thus, besides establishing as a unique instrument since the beginning of its post-commissioning life, the SKAO is due to also improve its capability in a relatively short timescale.

\section{The case of the SKA Regional Center Network (SRCnet)}

As said above, $\sim$ 700 PB of Observatory Data Products will be produced by SKAO in a year. However, in most cases they will not be ready yet for a full analysis and for the subsequent papers' writing by the scientists. Along the path already followed by other astrophysical projects, the Observatory Data Products will flow over the internet and will be distributed across a number of so-called SKA Regional Centers (SRCs), where they will be archived, cured and additionally manipulated by the scientists in order to produce the so-called Advanced Data Products (ADP). 

In view of this scheme, on July 2016 the Board of SKA (at the time provisionally leading the project) deliberated that {\it The SKA Observatory will coordinate a network of SKA Regional Centres (SRC) that will provide the data access, data analysis, data archive and user support interfaces with the user community}. To make the story short, during the subsequent years, a lot of in-kind efforts were done by many SKA members (including Italy) to collect the requirements from the science and software engineers community and to draft alternate possible design solutions for the network of the aforementioned SRCs, until March 2024, when the SKA IGO Council (the inter-governmental body which drives the project since February 2021) reviewed and finally endorsed the implementation of the network, dubbed as SRCnet. Shortly after, Rosie Bolton has been selected by the SKAO Director as the Interim SRCnet Project Lead. An SRCnet Resource board (mainly devoted to pledge human and hardware resources to the project and including one representative for each contributing Country) has also been set up, while the implementation of an Advisory Committee for the SRCnet Project is in progress. That will also include one representative for each contributing Country, plus other experts nominated by the Project Lead. Since mid 2022, the organization of the FTEs made available by the contributing Countries (in-kind at the beginning, in a formal pledged form hereafter) was done in the context of a series of teams, operating under the rules of the so-called Agile framework \cite{ref_article2}.

In summary, the SKAO and the SRCnet will be jointly responsible for: {\it (i)} maximizing the quality of SKA data delivered to users; {\it (ii)} 
generating the Advanced Data Products; {\it (iii)} storing, archiving curing the primary SKA output data and the Advanced Data Products; {\it (iv)} ensuring that the approved science program can be accommodated within available resources; {\it (v)} ensuring the availability of a platform of distributed services across computational and data infrastructures to support the user community to deliver SKA science, under the FAIR (Findable, Accessible, Interoperable, Reusable) principles.

The development will take place through a series of milestones, tailored to match the progress of the SKA construction, and thus aimed to a fully deployed system by 2027 at the latest. The first important step will be the set up of the first version of the network, v0.1, to be only used by the SRCnet project team as a test. That is expected to be progressively delivered during 2025. The interaction with the users will occur later, presumably towards the end of 2026, in coincidence with the parallel possibility for the users to inspect the first commissioning data produced by a small portion of the two instruments.

However, at that time the SRCnet will still need only a very limited amount of pledged hardware capabilities. The bulk of them should be made available towards the end of the 2027 and mid 2028, reaching $30-40$ Petaflop/s and $800-1000$ PB per year. In contrast with that trend, the needed investement in human capital should be done almost immediately, reaching $60-70$ FTE by the end of 2026 and then stabilizes at that level. 

\section{The Italian node of the SRCnet}

As reported above, Italy, via the National Institute for Astrophysics (INAF) has been an active member in the design of the SRCnet since the beginning and has given a significant in-kind contribution, fluctuating between 1 FTE and more than 2 FTEs, along all the trimesters of the Program Increments (PIs) occurred since mid 2022. In particular the work of the INAF team concentrated within one of the Agile teams, the so-called {\bf Orange team}, originally devoted to develop the tools of visualization of the radio astronomical data. In particular, the Orange team 
\begin{itemize}
\item contributed to the definition of the visualization use cases for SRCnet
\item performed a review of the Visualization Tools (dependencies, interfaces, workshop)
\item collected data products and data formats from precursors and pathfinders of SKA
\item adapted Visualization Tools to address use cases and work with SRC architecture and its data lake
\item developed, tested and deployed SODA (Server-side Operations for Data Access) into the SRCnet, integrated with Rucio Data Lake and Discovery services
\item reviewed Solutions and Technologies for the Computing Services API
\item tested and deployed the visualization tools and data access services into the SRCnet nodes
\item performed test of the performances of SODA, adapted SODA to work with storage manager service (SRM)
\item adapted SODA so that it can process requests through the Computing API (Application Programming Interface)
\item adapted the code VisIVO to invoke cutout and visualize data through the Computing API 
\end{itemize}  

Additional work was performed by INAF members also in other Agile teams, looking on one side at the complicated question of receiving and promptly storing the data coming from the 2 instruments, and, on another side, investigating about the design and implementation of an API to discover computing services implemented in the SRCnet, in turn answering the following typical questions for the users: {\it (a) what computing services are available and suitable in the SRCnet in order to run my task?} and {\it how can I run my task on the selected computing service (and when)?} 

\subsection{The plans for the future} 

At the first official pledging call for resources for the SRCnet, occurred in July 2024, INAF answered with about 3 FTEs, the largest level of involvement so far. On the other hand, that is not enough to satisfy the aim of INAF of at least matching the percentage of Italy in the construction budget, i.e. about 7\%. It is then foreseen the acquisition of new personnel only devoted to the maintenance and development of the Italian node of the SRCnet. The target should be to reach 6 FTEs by the end of 2026. 

As far as the hardware, INAF is completing the acquisition of a new system (provisionally called Tier-3) and, via the participation to the national center for computing ICSC (one of the large Infrastructures funded with the budget of the PNRR), 
INAF will also have priority access to many of the resources of the new Tier-1 run by the Cineca. In fact, both the Tier-1 and the Tier-3 will be localized at the Bologna Technopolo, where also the European weather centre ECMWF, the Leonardo super-computer, and the United Nations University on Climate Change are already, or will be hosted.

In this framework, the hardware resources made available by INAF for the v0.1 of the SRCnet might presumably be the following. As to computing power: 0.1 PFlop/s (Tier 3 – dedicated, with CPU only) by the beginning of 2025, and 1.5 PFlop/s (Tier 3 – dedicated, CPU+GPU) by the end of 2025, when also 15 additional PFlop/s (belonging to the Tier 1 – and as such to be used in a shared fashion, CPU+GPU) will become available. As to the storage, at the beginning of 2025, INAF will put online 0.3 PB on disks, and the equivalent of 1.2 PB on tapes. At the end 2025, the PB on disk will raise to about 2, whereas those on tapes (or similar devices tailored for long-term archiving) could reach 5 PB, plus about 10 PB of memory Flash (alike LUSTRE, shared).

A $2^{\rm nd}$ step in the grow of the INAF pledged hardware should occur by late 2027 or mid 2028, when a Tier-2/Tier-1.5 dedicated system with capability of $\sim 2$ Pflops and $\sim$ 60 PB of storage (20 PB on-line and 40 PB near on-line), will be connected at 100 Gb/s with the other nodes (v 1.0 of the SRCnet).

In conclusion, the Italian expected outcome from the SRCnet can be summarized as below:                   

\begin{itemize}
\item[1.] The identification of a kernel of {\it modi operandi} in the interactions among the various actors to secure an efficient, persistent, and always developable science-needs driven node 
\item[2.] The establishment of a node of the SRCnet  located in Italy
\item[3.] The recognition of the local investments in both hardware and human expertise, and its conversion into incentives as soon as possible 
\end{itemize} 

\begin{credits}
\subsubsection{\ackname} The author thanks the members of the now-expired SRC Steering Committee for the great ideas and support during the development of the concepts of the SRCnet. The author also acknowledges the contribution of the INAF Scientific Directorate in the development of the Italian node of the SRCnet.

\subsubsection{\discintname}
The author has no competing interests to declare that are
relevant to the content of this article. 
\end{credits}

\end{document}